# Demonstration of a Mid-infrared silicon Raman amplifier

**Varun Raghunathan[1], David Borlaug[1], Robert R. Rice[2] and Bahram Jalali[1]**

[1] *Electrical Engineering Department, UCLA, 420 Westwood Plaza, Los Angeles, CA 90095-1594, USA;*
[2] *Northrop Grumman Space Technology, One Space Park, Redondo Beach, CA 90278, USA*
*jalali@ucla.edu*

**Abstract:** We demonstrate, for the first time, a mid infrared silicon Raman amplifier. Amplification of 12 dB is reported for a signal at 3.39 micron wavelength. The active medium was a 2.5 cm long silicon sample that was pumped with 5ns pulses at 2.88 micron. The absence of the nonlinear losses, which severely limit the performance of silicon Raman devices in the near infrared, combined with unsurpassed crystal quality, high thermal conductivity render silicon a very attractive Raman medium. Such a technology can potentially extend silicon photonics' application beyond data communication in the near IR and into the mid IR world of remote sensing, biochemical detection and laser medicine.

©2007 Optical Society of America

## 1. Introduction

Silicon Photonics in the near infrared wavelength region has attracted significant attention in recent years with the aim of realizing low cost, high speed optoelectronic components for data and tele-communication applications. The enormous infrastructure available for silicon device manufacturing and the economy of scales have been the motivating factors in this direction. In particular, Raman scattering has been successfully used to demonstrate lasers [1,2] and amplifiers [3,4] on a silicon chip. One of the serious obstacles in demonstrating these devices in the near infrared has been overcoming the nonlinear loss mechanisms due to free-carrier and two-photon absorption. Low duty cycle pulse pumping [1,3] and reverse-biased carrier sweep out [2,4] were employed to minimize the impact of nonlinear losses. However, the maximum output power of these Raman devices is still limited by the nonlinear losses and the presence of free carriers also increases the on-chip heat dissipation [5].

Mid Wave Infrared (MWIR), defined loosely as the wavelength range spanning 2-6 micron, represents another set of applications were silicon photonics can play a role [6,7]. It has been experimentally shown that nonlinear absorption effects can be reduced to negligible levels by going to longer photon wavelengths for which the combined energy of two-photons is less than the band-gap of silicon [8]. The lack of nonlinear losses in the MWIR opens up new applications for silicon devices based on Raman and other nonlinear interactions. MWIR light sources and amplifiers are useful in applications such as free space communications, bio-chemical detection and certain medical procedures [9,10]. Most organic chemicals and biological agents have unique signatures in the MWIR and can be detected using these lasers. The strong water absorption peak at 2.9µm and the resulting applications in medicine and dentistry creates a large demand for such lasers. Typical laser sources developed for this wavelength range include diode pumped optical parametric oscillators [11], doped solid state lasers [12] and solid state (non semiconductor) Raman sources [13,14]. However, these sources have not achieved wide spread use owing to the complexity, high cost of implementation and poor reliability. Moreover, there is a need for efficient and versatile amplifiers in the MWIR to be used in power-scalable master oscillator power amplifier (MOPA) systems. Another new application is the recently proposed image pre-amplifiers [15]. Such a device employs Raman amplification combined with Talbot self imaging in a multimode waveguide. By amplifying the image above the thermal noise level of the optoelectronic image sensor, it promises to increase the sensitivity of MWIR remote sensing systems, or to allow them to operate without cooling. The missing prerequisite for experimental realization of such a device is Raman amplification in the MWIR, the very topic of the present paper.

In this paper we report on the first demonstration of a MWIR Silicon Raman amplifier. Specifically, an input signal at 3.39 micron wavelength is amplified by 12 dB in a 2.5cm long silicon sample. The sample was pumped with 5 ns long pump pulses at 2.88 micron wavelength. The paper starts with a brief description of the theory of Raman and other associated nonlinear optical effects with relevance to wavelength scaling. This is followed by the experimental set-up, results, and discussions.

## 2. Wavelength scaling of nonlinear optical effects

### A. Stimulated Raman scattering:

The Stimulated Raman process can be described as the interaction of the incoming pump and Stokes photons which sustains the material vibrations in the medium to efficiently stimulate the creation of Stokes photons. A detailed description of the process can be found in ref [16]. The macroscopic polarization of the stimulated scattering process is described in terms of the Polarization and a macroscopic third order Raman susceptibility [17]:

$$P_i^{(3)}(\omega_S) = \varepsilon_o \chi_{ijkl}^{(3)} E_j(\omega_S) E_k(\omega_p) E_l(-\omega_p) \tag{2.1}$$

In the case of single-crystal silicon, the Raman susceptibility has a typical resonant Lorentzian profile which peaks at the Raman shift (15.6 THz).

In the un-depleted pump approximation the evolution of the Stokes intensity along the propagation direction is conveniently described as: $\frac{dI_S}{dz} = g_R I_P I_S$, where the gain coefficient, $g_R$ is determined from the value of the Raman susceptibility $\chi^{(3)}$:

$$g_R = \frac{3\omega_S \mu_o}{n_S n_P} \chi_{ijkl}^{(3)} = \frac{6\pi\mu_o}{\lambda_S n_S n_P} \chi_{ijkl}^{(3)} \tag{2.2}$$

The gain coefficient extracted from stimulated Raman measurements in the near infrared (1550nm) is in the range of 10-20cm/GW [18,4]. For application to MWIR, the gain coefficient scales with wavelength as $\frac{1}{\lambda_S}$, as evident in the above equation (2.2). Thus, at the typical stokes wavelength used in our experiments of 3.39microns, the Raman gain coefficient is expected to be ~4.5-9cm/GW. In Raman amplification experiments in bulk crystals it has been found that the Raman gain scales down faster than the inverse wavelength scaling as predicted by theory [14].

### B. Nonlinear absorption:

The nonlinear absorption processes such as two-photon absorption (TPA) and free-carrier absorption (FCA) have been found to create additional loss mechanisms for the pump and Stokes photons and hence reduce the efficiency of the Raman process [19]. A schematic of the two-photon absorption process is shown in the inset of figure 1. TPA has been shown to be negligible from the point of view of pump depletion [18]. This is plausible since the TPA coefficient in silicon, β, is relatively small, ~ 0.5 cm/GW when compared to the Raman gain coefficient of $g_R$ = 20cm/GW at 1550nm wavelength. On the other hand, absorption by TPA-generated free carriers is a broadband process that competes with Raman gain. The magnitude of TPA induced free carrier absorption depends on free carrier concentration through the relation [20]:

$$\alpha^{FCA} = 1.45 \times 10^{-17} (\lambda/1.55)^2 \cdot \Delta N = \sigma \cdot \Delta N \tag{2.3}$$

here λ is the wavelength of interest (in microns), and ΔN is the density of electron-hole pairs. The latter is related to the pump intensity $I_p$, by:

$$\Delta N = (\beta_2 \cdot I_p^2 + 2\beta_2 \cdot I_p I_S) \cdot \tau_{eff} / (2 \cdot h\nu) \tag{2.4}$$

here $h\nu$ is the pump photon energy, and, $\tau_{eff}$, is the effective recombination lifetime for free carriers. This nonlinear absorption process co-exists with the Raman process and limits the maximum achievable amplification, beyond which loss overcomes the gain [20]. A lot of emphasis has been placed on minimizing the carrier concentration the near infrared region in-order to make the Raman process more efficient [1-4].

It has been experimentally shown that the two-photon absorption process diminishes in magnitude for photons with energy less than half the bandgap [8]. This is illustrated experimentally in figure 1 (reproduced from ref. [8]). As depicted in the inset of figure 1, photons at 2.09μm experience enough nonlinear absorption to create a nonlinear decrease in transmission of the incoming pump photons. On the other hand, in crossing half the bandgap by using pump photons of 2.936μm wavelength, the nonlinear process is completely eliminated. Thus the two-photon and free-carrier absorption processes reduce to negligible levels for photons of λ> 2.2μm (which corresponds to half the band-gap in silicon). Higher order nonlinear process such as three-photon absorption and the associated free-carrier absorption are expected to be insignificant because of the low probability of simultaneous absorption of three or more photons. Thus the TPA and associated FCA effects can be eliminated by employing mid infrared wavelengths (λ > 2.2μm) and hence the full potential of a Raman scattering can be realized.

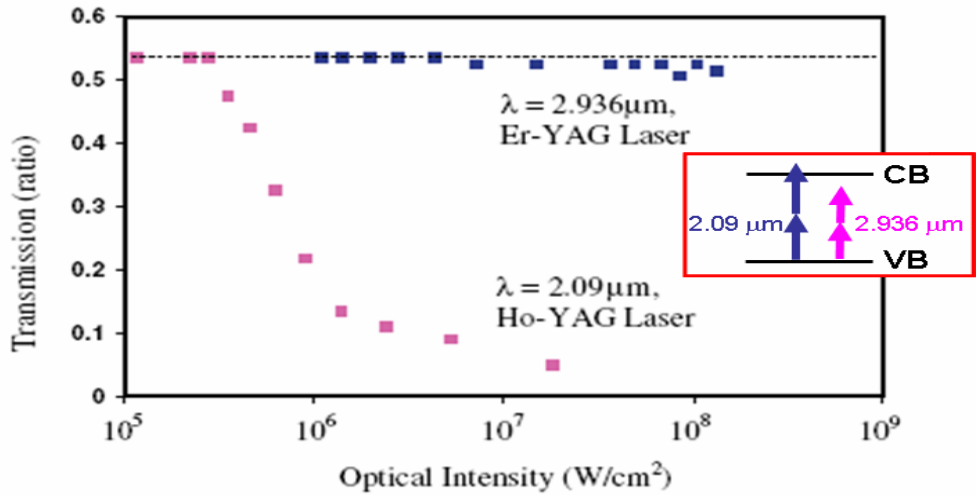

Figure 1: Transmission in silicon as a function of optical intensity. Two different pump sources at 2.09μm and 2.936μm were used in these experiments. The enhanced nonlinear losses at 2.09μm due to TPA and FCA and the absence of these losses at 2.936μm are clearly seen (figure reproduced from ref [8]).

## 3. Experiments

The experimental set-up used to observe Raman amplification in silicon in the MWIR wavelength region is shown in figure 2. The pump and Stokes signal lasers experiment were at 2.88μm and 3.39μm respectively. The pump laser was an optical parametric oscillator (OPO) and operated under pulsed condition with a pulse width of 5nsec and repetition rate of 10Hz. The source for the Stokes signal was a HeNe laser and operated under continuous-wave (CW) condition. The beam qualities of the pump and Stokes beams are $M^2 = 32$ and 2 respectively. At the input end, a dichoric beam-combiner was used to efficiently transmit the pump and reflect the Stokes beam and hence combine the two beams. A single plano-convex (PCX) lens was used to focus both the pump and Stokes laser into the silicon sample. The focal spot radius of the pump and Stokes beam were ~430μm and 120μm respectively. The poor beam quality of the pump ($M^2 = 32$) limited the focusing of the pump beam and hence the overlap of the pump and Stokes beams inside the silicon sample. The overlap between the pump and the Stokes beam inside the silicon sample was estimated to be ~14%.

The two beams were coupled into bulk [1 0 0] silicon sample of 2.5cm length. The two facets of this sample were coated with a broadband anti-reflection coating to prevent incurring fresnel reflection losses. At the output end a single $CaF_2$ plano-convex lens was used as the imaging lens. Two di-chroic beam-splitters were used to separate the strong residual pump from the weak amplified Stokes signal. A spectrometer was used to further filter the residual pump and observe the amplification at the Stokes wavelength. The time-resolved Stokes signal was detected using a cooled Indium Antimonide (InSb) detector and observed using an oscilloscope.

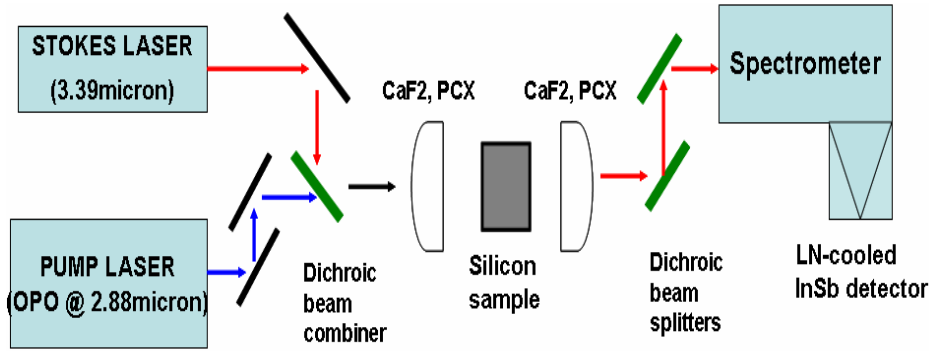

Figure 2: The experimental set-up used to observe Raman amplification in silicon at MWIR wavelengths. The pulsed pump OPO (2.88microns) and the CW Stokes HeNe laser (3.39microns) are focused and coupled into 2.5cm long AR-coated silicon sample with suitable $CaF_2$ plano-convex (PCX) lenses. At the output end, dichroic filters are used to block the residual pump and look at the amplified stokes using a fast InSb detector placed after a spectrometer.

Figure 3 shows the time-resolved Raman amplification results (averaged over 64 pulses). The Stokes laser polarization was fixed horizontal and the pump beam polarization was varied with a polarizer until the Raman gain signal in the oscilloscope is maximized. The typical on-off Raman gain obtained in the silicon sample was ~12dB at pump energy setting of ~3.5mJ over 5nsec. To the best of our knowledge, this is the first demonstration of optical amplification in silicon in the mid-wave infrared (MWIR) spectrum. It's important to note that the pump pulse width used here is only limited by the available pump laser and not by free carrier absorption effects in the gain medium [1,3]. The absence of free carrier absorption in the MWIR, as shown in Figure 1, suggests that there will be no limit on the maximum pulse width and continuous-wave amplification and lasing will be possible.

Figure 4 shows the plot of on-off Raman gain obtained as a function of the effective pump intensity. The apparent saturation of the Raman gain at higher pump intensities may be due to damage to the sample surface. This claim is supported by the results in figure 5 which shows the time-resolved Raman gain trace as the pump energy is increased. The d.c. level which corresponds to the input Stokes signal drops as the pump energy is increased. This is due to increase in linear absorption in the sample. The recovery of the dc level upon moving to a different spot on the silicon sample further supports this hypothesis. The damage may be caused in the AR coating, however, a detailed determination of the mechanism require further investigation and is beyond the scope of this paper.

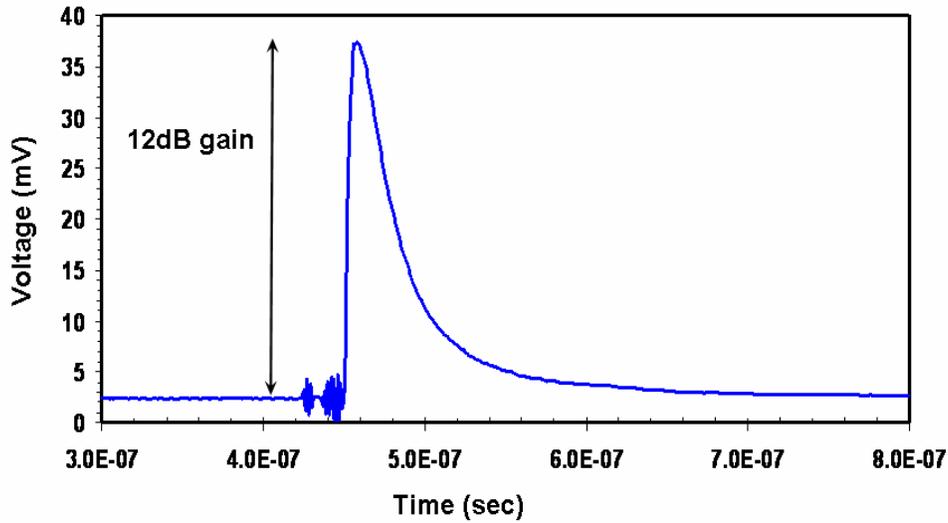

Figure 3: The time-resolved Raman amplification plot as observed using an oscilloscope. The typical Raman gain (on-off) obtained was ~12dB. The measurement was performed using a slow detector (~25nsec) when compared to the pump pulse (~5nsec) and hence the gain is underestimated due to smoothening of the trace. Trailing edge looks smoothened due to averaging. Averages of 64 traces were taken for this measurement.

The pump intensity giving rise to Raman amplification, after taking into account the overlap with the Stokes beam (~14%) is estimated to be 217MW/cm$^2$. Typical Stokes signal input of ~1-2mW was coupled into the silicon sample and such low signal levels ensured un-depleted pump condition. Taking into account the pump intensity calculated above and the Raman gain coefficient range of ~4.5-9 cm/GW (from section 2), the expected Raman gain expected is in the range 10.5-21 dB. The observed gain of 12 dB is towards the lower limit of this range. One reason is that the response time of the detector used to measure the Stokes gain was ~25nsec and was slower than the 5 ns pump pulse width. By smoothing out the Stokes pulse and not being able to capture its peak power, the detector underestimates the peak Raman gain.

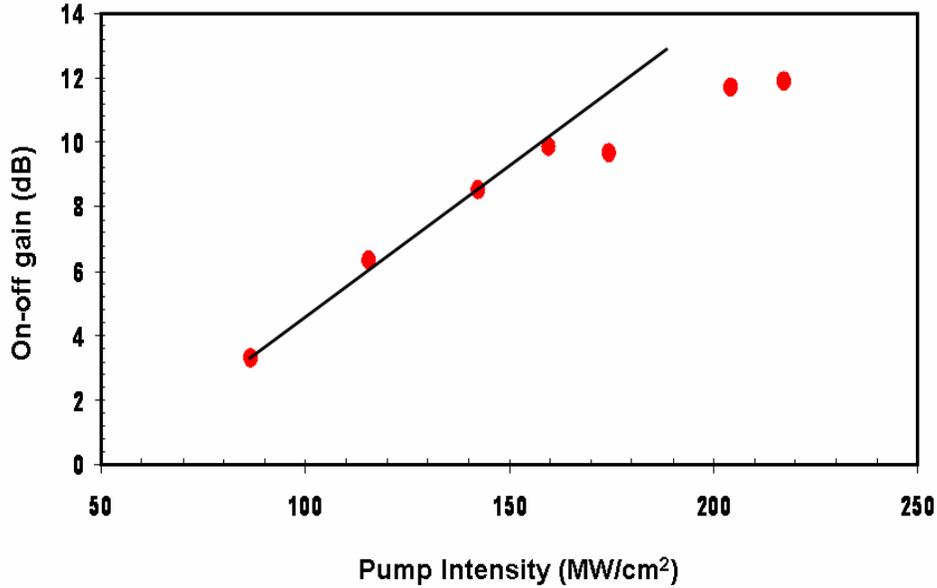

Figure 4: The plot of on-off Raman gain as a function of effective pump intensity interacting with the Stokes input. Each point is obtained by averaging 64 times. Maximum gain of 12dB is obtained at ~217MW/cm$^2$ pump intensity. A line is shown on this curve as a guide to the eye. The saturation seen in the curve is believed to be due to silicon damage.

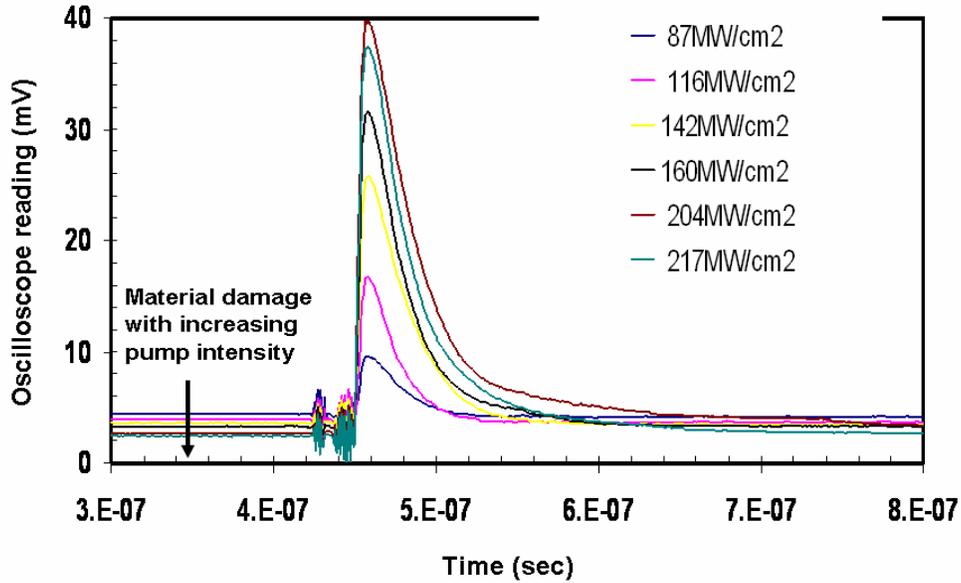

Figure 5: The plot of various time resolved Raman gain traces shown at increasing pump intensities. The drop in the dc level of the Stokes signal with increasing pump intensity is believed to be due damage of the silicon sample which increases the linear absorption of the sample.

## 4. Conclusions

In this paper we have demonstrated, for the first time, Raman amplification in silicon in the mid-wave infrared region of the spectrum. The mid infrared wavelength region is an appropriate window to build efficient silicon Raman devices owing to the absence of nonlinear absorption caused by two-photon and free-carrier absorption. This demonstration, combined with the well known low-loss and high thermal conductivity of silicon, underscores the case for mid-IR sources and amplifiers that are based on the silicon photonics technology.

**Acknowledgements:**
The authors would like to thank Prof. Oscar Stafsudd for helpful technical discussions. This work was supported by DARPA and Northrop Grumman Corporation.